\documentclass[twocolappendix]{emulateapj}

\usepackage[usenames,dvipsnames]{color}
\usepackage{comment}

\shorttitle{Analytical Gap Model}
\shortauthors{Paul Duffell}

\begin{document}

\title{A Simple Analytical Model for Gaps in Protoplanetary Disks}

\author{Paul C. Duffell}
\affil{Astronomy Department and Theoretical Astrophysics Center, University of California, Berkeley}
\email{duffell@berkeley.edu}

\begin{abstract}

An analytical model is presented for calculating the surface density as a function of radius $\Sigma(r)$ in protoplanetary disks in which a planet has opened a gap.  This model is also applicable to circumbinary disks with extreme binary mass ratios.  The gap profile can be solved for algebraically, without performing any numerical integrals.  In contrast with previous one-dimensional gap models, this model correctly predicts that low-mass (sub-Jupiter) planets can open gaps in sufficiently low-viscosity disks, and it correctly recovers the power-law dependence of gap depth on planet-to-star mass ratio $q$, disk aspect ratio $h/r$, and dimensionless viscosity $\alpha$ found in previous numerical studies.  Analytical gap profiles are compared with numerical calculations over a range of parameter space in $q$, $h/r$, and $\alpha$, demonstrating accurate reproduction of the ``partial gap" regime, and general agreement over a wide range of parameter space.

\end{abstract}


\section{Introduction} \label{sec:intro}

Disk-satellite interactions have been studied for decades.  The general problem of an orbiting point mass interacting gravitationally with a surrounding disk of either gas or solids is a fundamental physics problem with a multitude of applications in astrophysics.  Saturn's rings, for example, are an exquisite testbed for disk-satellite interactions in the case where the disk is composed entirely of solids \citep{1978Icar...34..227G}.  For the hydrodynamic case, how a planet interacts with the gaseous disk which spawned it is of great importance to the question of how the planets formed.  It also provides a vital keystone in our interpretation of observational data, both of extrasolar planetary systems and of debris disks around newly formed stars.  Additionally, disk-satellite interactions are of importance to the orbital evolution of a black hole binary embedded in a circumbinary disk.

The linear theory of hydrodynamical disk-planet interactions has been very successful so far \citep{1978Icar...34..227G, 1980ApJ...241..425G, 1986Icar...67..164W}.  Predictions of the disk's linear response to the planet and subsequent predictions of the disk perturbation's gravitational influence on the planet have been accurately reproduced in numerical studies.  This is true for questions of planetary (Type I) migration rates \citep[e.g.][]{2002ApJ...565.1257T, 2010ApJ...724..730D}, and of the eccentricity evolution of the planet \citep[e.g.][]{2006AnA...450..833C, 2007AnA...473..329C, 2010AnA...523A..30B}.

The details of nonlinear theory, however, have not all been reproduced in numerical studies.  In linear theory, a spiral wave is excited by the planet, and passes politely through the disk.  After each wave front passes, it leaves no trace of its presence.  A nonlinear wave, however, can shock.  When it does so, it imparts its angular momentum to the disk, causing fluid elements to move away from the planet's position.  This process carves out a low-density annulus in the vicinity of the planet's orbit.  This low-density region is known as a ``gap", and its detailed properties have eluded a robust and numerically reproducible semi-analytical description for some time.

Many attempts have been made to describe this three-dimensional system with a one-dimensional model, to predict the gap structure, i.e. the surface density $\Sigma$ as a function of radius $r$.  \cite{2004ApJ...612.1152V} took the disk evolution equations of \cite{1974MNRAS.168..603L} and added a planetary torque term, to predict the width and shape of gaps.  The calculated gaps turned out to be much wider than those seen in numerical studies.  \cite{2006Icar..181..587C} attempted to fix this by adding a component due to ``pressure torque" in the disk.  However, both of these studies predicted gaps which were much too deep.  In part, this is because the model for torque deposition was too simplistic; for example, the torque from the planet was calculated locally, even though the transfer of angular momentum from the planet to the disk by shocks is an inherently nonlocal process, as the wave must propagate some distance before shocking.  Additionally, these studies considered the role of planetary torque as a barrier holding back the viscous accretion of gas.  Such considerations predict strict criteria for gap opening \citep{1979MNRAS.186..799L, 1993prpl.conf..749L, 1999ApJ...514..344B, 2006Icar..181..587C} but semi-analytical and numerical studies have demonstrated that planets can violate these criteria and still open gaps on long timescales \citep{2001ApJ...552..793G, 2011ApJ...741...56D, 2011ApJ...741...57D, 2012ApJ...755....7D}.

Similarly problematic assumptions have been made for 1D models of circumbinary disks.  Disk models of \cite{1995MNRAS.277..758S} and later \cite{2012MNRAS.427.2660K} implicitly assume that if the planet does not migrate with the gas, there is a ``pile-up" of gas at the outer edge of the gap.  In contrast, the disk model of \cite{2006ApJ...641..526L} considered gas flow across the gap, but even that study assumed that the planet's presence affected the accretion rate at infinity.

Not only are many of these assumptions incorrect \citep[see for example][]{2014ApJ...792L..10D}, but they also result in gap profiles which are inconsistent with numerical experiments.  The gap profiles in these studies generally predict exponentially deep gaps; that is, the surface density near the planet is depleted by a factor $\sim {\rm exp}(-(q/q_{\rm gap})^2)$ relative to the unperturbed state, where $q$ is the planet-to-star mass ratio, and $q_{\rm gap}$ is some critical threshold for gap opening.  On the other hand, several recent numerical parameter surveys have shown that gap depth scales as a power-law with $q$, $h/r$ and $\alpha$ \citep{2013ApJ...769...41D, 2014ApJ...782...88F}, so that gaps are not as depleted as in the 1D models.

\cite{2014ApJ...782...88F} provided an argument for the power-law scalings found in \cite{2013ApJ...769...41D}, pointing out that the torque produced by the planet would be proportional to the surface density in the gap near the planet's position $\Sigma_p$, assuming that the torque was dominated by resonances excited inside the gap.  Very recently, \cite{2015MNRAS.448..994K} produced a 1D model for gaps which repeated the arguments of \cite{2014ApJ...782...88F}, recovering the power-law scaling of \cite{2013ApJ...769...41D}.

However, \cite{2015MNRAS.448..994K} used an incomplete description of the planetary torque.  For most of their study, the torque was specified as the \emph{excitation} torque in the disk, as calculated by linear theory, whereas the torque which controls gap opening is the \emph{deposition} torque, i.e. the rate of transfer of angular momentum from the planetary wake to the disk by shocks.  They attempted to fold this into their study by including a parameter $x_d$ which measured the distance the wave travels from the planet before shocking.  The resultant gap profile was then some function of the quantity $x_d$, which was treated as a free parameter.

Fortunately, the process of excitation and deposition of torque in the disk has been studied by \cite{2001ApJ...552..793G} (hereafter GR01).  Therefore, there is no need to speculate about this process, at least in the weakly nonlinear regime.  In the present study, the planetary angular momentum flux derived by GR01 is applied to a disk model similar to \cite{2015MNRAS.448..994K}.  These disk equations can then be solved \emph{algebraically} for the surface density, so that it is not necessary to perform any numerical integrals as in \cite{2006Icar..181..587C} and \cite{2015MNRAS.448..994K}.

The model is described in section \ref{sec:anly}.  In section \ref{sec:results}, the resultant gap profile $\Sigma(r)$ is compared with gap profiles from a recent parameter survey \citep{2014arXiv1412.8092D} and agreement is found, particularly in the ``partial gap" regime.  This provides confidence that a complete 1D disk model may be possible after taking into account higher-order effects.  A summary and discussion is given in section \ref{sec:disc}.

These results may be testable observationally in the not-too-distant future, as the ALMA project may be capable of measuring surface density profiles in transition disks.

\section{Description of the Analytical Model} \label{sec:anly}

The initial approach here is essentially equivalent to that of \cite{2004ApJ...612.1152V}, though the underlying equations will be derived in a slightly different manner.  Rather than using the equation of mass flux to derive a differential equation for $\Sigma(r)$, a planetary torque term is inserted into the equation of angular momentum flux, and the fluxes at infinity are assumed to be unaffected by the presence of the planet.  The system is then solved algebraically.

The equations of mass and angular momentum conservation in a steady-state viscous disk are as follows (ignoring the planet, at first):

\begin{eqnarray}
\dot M = - 2 \pi r \Sigma v_r = \text{const}\\
\dot J = - \dot M r^2 \Omega - 2 \pi r^3 \Sigma \nu {d\Omega \over dr} = \text{const}
\label{eqn:jdot}
\end{eqnarray}

where $\dot M$ is the mass accretion rate, $\dot J$ is the angular momentum flux, $v_r$ is the radial drift velocity, $\Omega$ is the orbital angular velocity of the gas, and $\nu$ is the kinematic viscosity.  All quantities are vertically and azimuthally averaged.  The accretion rate is defined so that positive $\dot M$ implies inward accretion.  The first term in (\ref{eqn:jdot}) is the flux due to advection, and the second term is due to viscous torques.

Assuming Keplerian $\Omega$ (as will generally be done for simplicity in this study), equation (\ref{eqn:jdot}) becomes
\begin{equation}
\dot J = - \dot M r^2 \Omega + 3 \pi \Sigma \nu r^2 \Omega.
\end{equation}

There are two possible solutions such that $\dot M$ and $\dot J$ are both spatially uniform.  The first is $\dot M = 0$, $\dot J = 3 \pi \Sigma \nu r^2 \Omega = $ constant, and the second is $\dot M = 3 \pi \Sigma \nu = $ constant, $\dot J = 0$, so that viscous angular momentum transport outwards exactly cancels advective transport inwards.  The latter solution will be concentrated on in this study, as there is zero accretion in the former solution.  However, the $\dot M = 0$ solution does not pose any additional challenges, and what follows would be nearly identical (a linear combination of the two is also possible, but again this is ignored in the current study for simplicity).

Thus, for this (unperturbed) solution, the surface density $\Sigma$ is inversely proportional to the viscosity, $\nu$.  If $\nu$ is spatially uniform, then $\Sigma$ will also be, in steady state.

Once the constants $\dot M$ and $\dot J$ are known, one can add a planetary torque as an additional source of angular momentum flux:

\begin{equation}
\dot J = - \dot M r^2 \Omega + 3 \pi \Sigma \nu r^2 \Omega + \Phi_p(r) = 0
\label{eqn:phip}
\end{equation}

$\Phi_p(r)$ is the angular momentum flux generated by the planet; the torque density deposited by shocks is given by the negative derivative of this function.

Note that it is assumed that the planet does not affect the values of $\dot M$ and $\dot J$ at infinity, so that in particular the gap does not act as a ``barrier" preventing mass flux through the system.  The surface density $\Sigma(r)$ adjusts itself to maintain these values of $\dot M$ and $\dot J$ in the presence of the planet.  The value of $\dot M$ is the unperturbed value, the same as it would be in absence of the planet:

\begin{equation}
\dot M = 3 \pi \nu \Sigma_0(r)
\label{eqn:mdot}
\end{equation}

(the viscosity $\nu$ can also depend on radius, but this notation was omitted for simplicity).  The planet's influence is entirely housed in the planetary angular momentum flux, $\Phi_p(r)$.  $\Phi_p(r)$ is determined by how far the wave propagates before shocking, and how quickly its angular momentum is transferred to the disk after shocking.  This process has been studied by \cite{2001ApJ...552..793G}, and their calculations provide a clear prescription for $\Phi_p(r)$ which applies in the weakly nonlinear regime.  The specific details of this function will be discussed further in section \ref{sec:phir}, but the overall scaling is given by the one-sided torque emanating from the planet:

\begin{equation}
\Phi_p(r) = \Sigma_p a^4 \Omega_p^2 q^2 \mathcal{M}^3 f(r)
\end{equation}

where $a$ is the radial position of the planet, $\Sigma_p$ is the surface density at the planet's position $\Sigma(a)$, $q$ is the planet-to-star mass ratio, and $\mathcal{M}$ is the Mach number (the inverse of the disk aspect ratio $h/r$).  $f(r)$ is a dimensionless function of $r$, discussed later.  Near the planet, $f$ evaluates to $f(a) = f_0$, where $f_0$ is some dimensionless order-unity constant.  Note that there is an implicit assumption that the waves are excited within the gap, so that the angular momentum flux is proportional to the density in the gap, $\Sigma_p$.

Evaluating (\ref{eqn:phip}) at the planetary position (also incorporating (\ref{eqn:mdot}) for the mass flux) gives:

\begin{equation}
- 3 \pi \nu \Sigma_0 a^2 \Omega_p + 3 \pi \Sigma_p \nu a^2 \Omega_p + \Sigma_p a^4 \Omega^2 q^2 \mathcal{M}^3 f_0 = 0.
\end{equation}

This can be inverted to find a formula for gap depth:

\begin{equation}
{\Sigma_p \over \Sigma_0} = ( 1 + f_0 {q^2 \mathcal{M}^3 a^2 \Omega_p \over 3 \pi \nu} )^{-1}.
\end{equation}

\begin{figure}
\epsscale{1.2}
\plotone{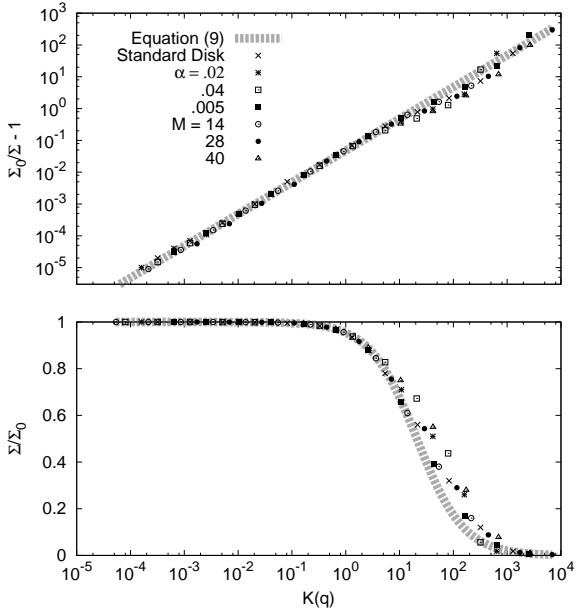}
\caption{ Gap depth $\Sigma_p / \Sigma_0$ is plotted for 84 different planet-disk models, from the parameter survey of \cite{2014arXiv1412.8092D}.  These depths are compared with equation (9).  Planet-to-star mass ratio ranged from $q = 10^{-6}$ to $q = 2 \times 10^{-3}$, Mach number ranged from $\mathcal{M} = 14$ to $\mathcal{M} = 40$, and viscosity ranged from $\alpha = 0.005$ to $\alpha = 0.04$.  The horizontal axis plots the quantity $K(q) = q^2 \mathcal{M}^5 / \alpha$.
\label{fig:depths} }
\end{figure}

This recovers the gap depth scaling reported in \cite{2013ApJ...769...41D}.  This exercise was also effectively carried out by \cite{2014ApJ...782...88F}, and by \cite{2015MNRAS.448..994K}.  More specifically,

\begin{equation}
{\Sigma_p \over \Sigma_0} = { 1 \over 1 + f_0 K(q) / (3 \pi) }
\label{eqn:A}
\end{equation}

where 

\begin{equation}
K(q) \equiv q^2 \mathcal{M}^5 / \alpha
\end{equation}

and $\alpha$ is the dimensionless Shakura-Sunyaev viscosity parameter, $\alpha = \nu \mathcal{M}^2 / (a^2 \Omega)$.  Figure \ref{fig:depths} compares gap depths from equation (\ref{eqn:A}) with steady-state 2D solutions from a numerical parameter survey over $q$, $\alpha$ and $\mathcal{M}$ \citep{2014arXiv1412.8092D}, demonstrating the accuracy of this formula over a wide range of parameter space.  For these 84 disk models, the only significant deviation from equation (\ref{eqn:A}) is in the $K \sim 10-1000$ range, where higher-order effects become important.  For larger values of $K$, \cite{2014ApJ...782...88F} found that this scaling eventually changes, suggesting that higher-order effects become important for very large $K > 10^4$.  This figure was used to constrain the one free parameter $f_0 = 0.45$.  This value of $f_0$ is also consistent with the one-sided torque calculated numerically in these disk models ($f_0 \approx 0.451 \pm 0.534/\mathcal{M}$, where the $\pm$ sign indicates the torque excited on either side of the planet, with the greater torque excited in the outer disk).  After choosing $f_0 = 0.45$, the model is completely specified.

\subsection{Density as a Function of Radius} \label{sec:phir}

Now with this value of $\Sigma_p$, one can return to the original $\dot J$ equation:

\begin{equation}
- 3 \pi \nu \Sigma_0(r) r^2 \Omega + 3 \pi \Sigma(r) \nu r^2 \Omega + \Sigma_p a^4 \Omega_p^2 q^2 \mathcal{M}^3 f(r) = 0.
\label{eqn:final}
\end{equation}

Given $\Sigma_p / \Sigma_0$, which was calculated above (\ref{eqn:A}), one can solve (\ref{eqn:final}) for $\Sigma(r)$:

\begin{equation}
\Sigma(r) = \Sigma_0(r) \left( 1 - {{ f(r) K(q) / (3 \pi) } \over 1 + f_0 K(q) / (3 \pi) } \sqrt{a/r} \right)
\label{eqn:B}
\end{equation}

The function $f(r)$ is a scaled-out version of the angular momentum flux due to the shocking of the planetary wake, which has been calculated semi-analytically by \cite{2001ApJ...552..793G} in shearing-box coordinates, and this result has been extended to global disk coordinates by \cite{2002ApJ...569..997R}.  Therefore, there is a well-specified formula for $\Phi_p(r)$ throughout the entire disk, specifically applicable for low-mass planets whose planetary influence is weakly nonlinear.  By design, this is only meant to apply if the quantity $q \mathcal{M}^3 \ll 1$, meaning for sub-Jupiter planets.  However, it may be possible to improve the semi-analytic calculation of $\Phi_p(r)$, leading to an accurate model for all planet masses.  This will be attempted in a future study.

For the present study, the formula for $\Phi_p(r)$ comes from the calculations of \cite{2001ApJ...552..793G} and \cite{2002ApJ...569..997R}.  The wave conserves its angular momentum until it shocks and deposits its angular momentum into the disk.  Therefore, the function $f(r)$ is simply a constant, until the distance from the planet that the wave shocks.  The explicit shape of this function is shown in Figure 3 of GR01, and its extension to a global disk is shown in Figure 3 of \cite{2002ApJ...569..997R}.  The function $f(r)$ can be described in terms of the parameter $\tau(r)$:

\begin{equation} 
f(r) = \left\{ \begin{array}
				{l@{\quad \quad}l}
				f_0 & \tau(r) < \tau_{sh} 	\\  
    			f_0 \sqrt{\tau_{sh}/\tau(r)} & \tau(r) > \tau_{sh} 	\\ 
    			\end{array} \right.    
\label{eqn:C}
\end{equation}

\begin{figure}
\epsscale{1.2}
\plotone{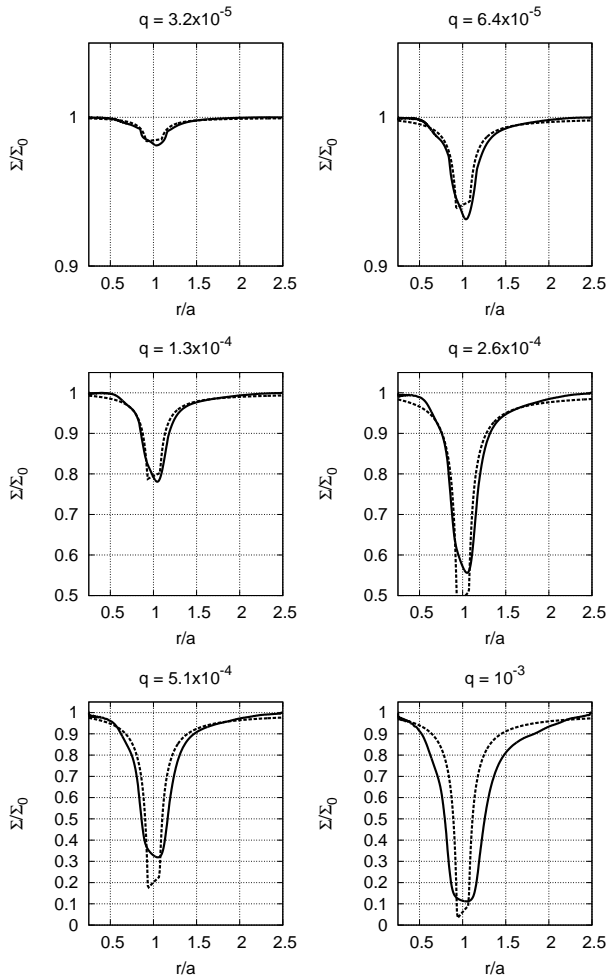}
\caption{ Several planet masses test the 1D model, for a disk with $\mathcal{M} = 20$, $\alpha = 0.01$.  Solid curves are numerical, while dashed curves are analytical.  Agreement is found for lower-mass planets, in the ``partial gap" regime.  For more massive planets ($q \sim 10^{-3}$, or Jupiter-mass), there is less agreement.  However, improvements may be made by taking into account the change in the torque due to the detailed gap structure.
\label{fig:mass} }
\end{figure}

where the shock position $\tau_{sh}$ was calculated by \cite{2001ApJ...552..793G}, and is given by

\begin{equation}
\tau_{sh} = 1.89 + 0.53 / (q \mathcal{M}^3)
\label{eqn:D}
\end{equation}

The parameter $\tau(r)$ represents an appropriately-scaled distance from the planet.  In shearing-box coordinates, this is given by

\begin{equation}
\tau(r) = {2^{3/4} \over 5} |{3 \over 2}\mathcal{M}(r/a-1)|^{5/2}
\end{equation}

To apply this to a global disk, in which surface density and sound speed vary as $\Sigma \propto r^{-p_1}$ and $c \propto r^{-p_2}$, one must perform the following integral

\begin{equation}
\tau(r) = {3 \over 2^{5/4}} \mathcal{M}^{5/2} \left| \int_1^{r/a} |s^{3/2} - 1|^{3/2} s^{(5p_2+p_1)/2-11/4} ds \right|
\label{eqn:E}
\end{equation}

which can be integrated analytically and expressed in terms of hypergeometric functions.

Equations (\ref{eqn:B}),(\ref{eqn:C}),(\ref{eqn:D}), and (\ref{eqn:E}) completely specify the value of the surface density at every point in the disk, given the parameters $q$, $\alpha$, and $h/r$, and the background surface density profile $\Sigma_0(r)$.

\section{Comparison with Numerical Results} \label{sec:results}

\begin{figure}
\epsscale{1.2}
\plotone{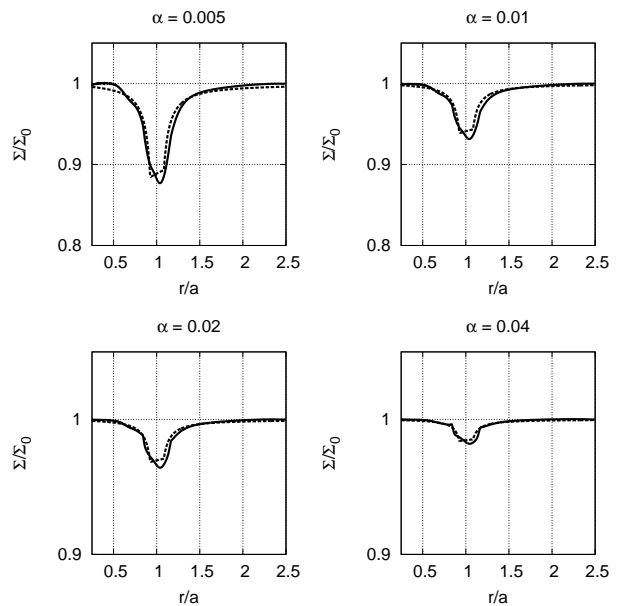}
\caption{ Variation with viscosity is tested for the $q = 6.4 \times 10^{-5}$ ($\sim 20 M_{\earth}$) planet, in a disk with $\mathcal{M} = 20$.
\label{fig:visc} }
\end{figure}

The accuracy and range of applicability of this simple analytical model are tested by comparing directly with numerical calculations from the extensive parameter survey of \cite{2014arXiv1412.8092D}.  In that survey, three parameters were varied: mass ratio $q$, Mach number $\mathcal{M}$ (or equivalently $h/r = 1/\mathcal{M}$), and viscosity $\nu$ (or equivalently, $\alpha = \nu \mathcal{M}^2 / (a^2 \Omega_p)$.

Figure \ref{fig:mass} shows variation with mass ratio for a disk with $\mathcal{M} = 20$ and $\alpha = 0.01$.  Mass ratios from $q = 3.2 \times 10^{-5}$ to $q = 10^{-3}$ are studied showing a wide range of applicability of the 1D model.  Near $q = 10^{-3}$ (Jupiter-mass), the model begins to break down as the gap is significantly wider than predicted, but the depth of the gap is still reasonably accurate.

The reason for the deviation around Jupiter-mass is likely that all of the excitation is assumed to come from within the gap, whereas once the gap becomes deep enough, the dominant resonances live just outside the gap edges.  It is likely that a great deal of improvement would come from modifying the function $\Phi_p(r)$ to account for excitation of waves in the presence of the detailed shape of the gap.

\begin{figure}
\epsscale{1.2}
\plotone{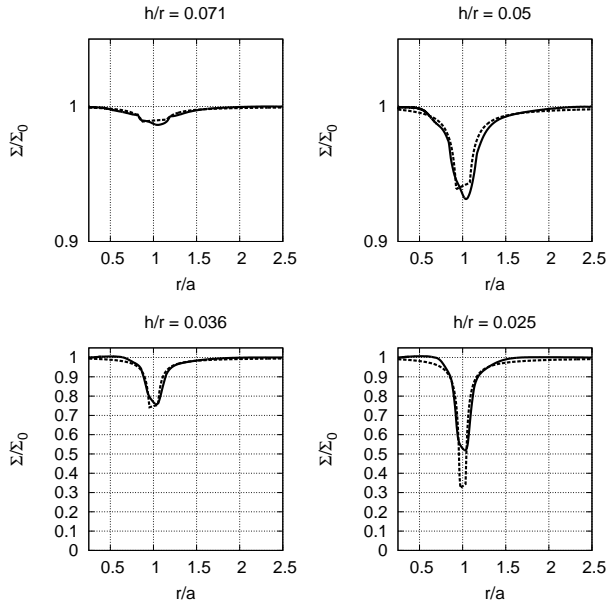}
\caption{ Dependence on disk temperature is explored, by varying the Mach number $\mathcal{M}$ (or equivalently $h/r = 1/\mathcal{M}$).  $\alpha$ is kept fixed, whereas $\nu = \alpha a^2 \Omega_p / \mathcal{M}^2$ is varied.
\label{fig:temp} }
\end{figure}

Figure \ref{fig:visc} shows variation of the gap profile with viscosity, for a $q = 6.4 \times 10^{-5}$ ($\sim 20 M_{\earth}$) planet in a disk with $h/r = 0.05$ ($\mathcal{M} = 20$).  The gap depth and shape is accurately captured for this range of viscosities.

Finally, Figure \ref{fig:temp} shows dependence on disk temperature (equivalently $\mathcal{M}$ or $h/r$) for $q = 6.4 \times 10^{-5}$, $\alpha = 0.01$ ($\alpha$ is fixed, but $\nu = \alpha a^2 \Omega_p / \mathcal{M}^2$ varies with $\mathcal{M}$).  The shape of the gap is reasonably well-captured, but again once the gap becomes deep enough, there is a similar deviation from the model (this same deviation occurs at around $K \sim 100$ in Figure \ref{fig:depths}; note that the last panel in this figure has $K = 42$).

This analytic model should be improved for the case of deep gaps (this is planned for a future study), but it accurately captures the ``partial gap" regime, and the basic fundamentals are correct in that the scalings of \cite{2013ApJ...769...41D} are reproduced, and low-mass planets can also open gaps in sufficiently low-viscosity disks, which has also been seen numerically \citep{2011ApJ...741...56D, 2011ApJ...741...57D, 2012ApJ...755....7D}.

\section{Discussion} \label{sec:disc}

A simple analytical model has been presented for gap profiles in protoplanetary disks.  In this model, effects due to planetary torque come directly from the detailed calculations of shock propagation by GR01, and \cite{2002ApJ...569..997R}.

The model is solved algebraically, and compared with long-duration numerical calculations.  The model's performance in the ``partial gap" regime is unprecedented in its agreement with numerical calculations and its reproduction of empirically found scalings.  For deep gaps, the model does not correctly predict the gap shape.  This is most likely due to the fact that resonances located several scale heights from the planet become dominant once the gap suppresses the resonances nearest the planet.  Additionally, effects due to non-Keplerian orbital motion should become important for very deep gaps.

One could envision an iterative procedure for folding these effects into the analytic formula.  For example, given the profile $\Sigma(r)$, one could calculate the contribution of the torque from all resonances in the disk.  This could give a new form for $\Phi_p(r)$, which would provide a corrected surface density via equation (\ref{eqn:final}).  This process could be repeated until a converged gap profile is found.  Such a procedure will be attempted in a future study.

\acknowledgments

Resources supporting this work were provided by the NASA High-End
Computing (HEC) Program through the NASA Advanced Supercomputing (NAS)
Division at Ames Research Center.

I am grateful to Eugene Chiang, Roman Rafikov, Andrew MacFadyen, Steve Stahler, Eliot Quataert, Kazuhiro Kanagawa, Alessandro Morbidelli, and Aurelien Crida for helpful comments.

\bibliographystyle{apj} 

\end{document}